\documentclass[aps,pra,letterpaper,10pt,twocolumn]{revtex4-1}
\usepackage[colorlinks=true,allcolors=blue]{hyperref}
\usepackage{amssymb}
\usepackage{amsmath}
\usepackage{graphicx}
\usepackage[caption=false]{subfig}
\usepackage{amsfonts}
\usepackage{xcolor}
\usepackage{epsfig}
\usepackage{color}
\usepackage{bm}
\usepackage{tabularx}
\usepackage{multirow}
\usepackage{mathtools}
\usepackage{xfrac}
\usepackage{blkarray}
\usepackage{bbold}
\usepackage[mathscr]{euscript}
\usepackage{autobreak}
\usepackage{comment}

\newcommand{\vncb}{V$_{\rm N}$C$_{\rm B}$}
\newcommand{\cnvb}{C$_{\rm N}$V$_{\rm B}$}
\newcommand{\cncb}{C$_{\rm N}$C$_{\rm B}$}
\newcommand{\vn}{V$_{\rm N}$}
\newcommand{\vb}{V$_{\rm B}$}
\newcommand{\cn}{C$_{\rm N}$}
\newcommand{\cb}{C$_{\rm B}$}

\begin{document}
\title{Bistable carbon-vacancy defects in $h$-BN}

\author{Song Li}
\affiliation{Wigner Research Centre for Physics, P.O.\ Box 49, H-1525 Budapest, Hungary}

\author{Adam Gali}
\affiliation{Wigner Research Centre for Physics, P.O.\ Box 49, H-1525 Budapest, Hungary}
\affiliation{Department of Atomic Physics, Institute of Physics, Budapest University of Technology and Economics, M\H{u}egyetem rakpart 3., H-1111 Budapest, Hungary}

\date{\today}
\begin{abstract}
Single photon emitters in hexagonal boron nitride have been extensively studied recently. Although unambiguous identification of the emitters is still under intense research, carbon related defects are believed to play a vital role for the emitter producing zero-phonon-lines in the range of $1.6$ to $2.2$~eV. In this study, we systematically investigate two configurations of carbon-vacancy defects, \vncb\ and \cnvb, by means of density functional theory calculations. We calculated the reaction barrier energies from one defect to the other to determine relative stability. We find the barrier energies are charge dependent and \cnvb\ could easily transform to \vncb\ in neutral and positive charge states while it is stable when negatively charged. Formation energy calculations show that the \vncb\ is the dominant defect over \cnvb.  However, neither \vncb\ nor \cnvb\ has suitable  fluorescence spectra that could reproduce the observed ones. Our results indicate that the origin of the $1.6$-to-$2.2$-eV emitters should be other carbon-related configurations.
\end{abstract}

\maketitle

%
%
\section{Introduction}
Point defects as emerging single photon emitters (SPEs) in two-dimensional (2D) hexagonal boron nitride (hBN) have been intensively studied for possible applications in quantum sensing, computing and nanophotonics~\citep{tran2016quantum,gottscholl2020initialization,chejanovsky2021single,mendelson2021identifying,hayee2020revealing,bourrellier2016bright,bommer2019new,tran2016robust}. The wide band gap ($\sim$6 eV) and small spin-orbital coupling of hBN manifests itself as ideal host to accommodate color centers in a wide region of emission wavelength. The spatial confinement and dielectric screening of 2D defects enable them showing desirable properties, for example, bright luminescence, ease of manipulation, tunable emission through strain~\citep{grosso2017tunable,hayee2020revealing,mendelson2020strain} and electric fields~\citep{noh2018stark} motivating researchers to investigate the underlying physical and chemical nature of the structures in detail. However, direct mapping or characterization of the defects in experiment is still a challenge. This might be partially related to various defect types and unintentional impurities during hBN sample fabrication and post-processing.

Photoluminescence data reveals that there are strong emission bands in ultraviolet region with zero-phonon-line (ZPL) energy at 4.1~eV~\citep{bourrellier2016bright,vuong2016phonon,du2015origin,museur2008defect} and in visible region from 1.6 to 2.2~eV~\citep{tran2016robust,sajid2020single,li2020giant,tran2016quantum,gottscholl2020initialization,chejanovsky2021single,mendelson2021identifying}. Based on this, many kinds of point defects have been proposed and analyzed theoretically through density functional theory (DFT) calculations. Several of them could match the experimental result well, such as the boron vacancy (\vb)~\citep{abdi2018color,ivady2020ab,reimers2020photoluminescence}, nitrogen vacancy (\vn)~\citep{sajid2018defect}, and Stone-Wales defect ~\citep{wang2016local,hamdi2020stone}. \vb\ have been identified experimentally~\citep{jin2009fabrication,gottscholl2021room,liu2021rabi,kianinia2020generation} and single spins of other defects could be coherently manipulated~\citep{chejanovsky2021single, stern2022room}. Besides, defect complexes and impurities that deliberately incorporated during synthesis or ion implantation also could serve as potential candidates~\citep{sajid2018defect,wu2019carrier,weston2018native,gottscholl2020initialization,chejanovsky2021single,mendelson2021identifying}. Among them, carbon impurity in hBN has been extensively considered and investigated. Previous study indicated that the recombination from \cn\ as a donor-acceptor pair with \vn\ donor~\citep{du2015origin} should not be related to 4.1-eV emission line due to the deep donor level of \vn\ whereas the \cb\ might be a possible source with charge transition level $(0/+)$ at 3.71~eV~\citep{weston2018native}. Later carbon dimer \cncb\ was also associated to this emission with a calculated ZPL at 4.3~eV~\citep{mackoit2019carbon,li2022ultraviolet}. For the visible emission, both \vncb\ and \cnvb\ are considered depending on their charge states and spin multiplicity. Our study already revealed the positively charged \vncb$(+)$ has ZPL about 1.5~eV which could be classified to so called “Group-2” SPE~\citep{abdi2018color}. In the triplet electronic configuration of its neutral charge state, the two defects have ZPL optical transitions at 1.58~eV and 1.54~eV, respectively~\citep{sajid2018defect}. A recent study has indicated that the quartet state of the negatively charged \cnvb$(-)$ is the origin of the SPEs in carbon doped hBN~\citep{mendelson2021identifying}.

The robust emission features from carbon defects in hBN samples grown with different techniques and environment indicate that they come from very stable configurations. However, considering the fact that the carbon atom in \vncb\ and \cnvb\ structures can migrate easily under external perturbation, it is necessary to investigate the relative stability and transformation of the two defects in detail. Formation energy and energy barrier calculation can help to identify the possible structures. The experimental observed ZPL emission is about 1.6 to 2.2~eV with Huang-Rhys factor less than 2. Here we use these criteria to examine the possibility of \vncb\ and \cnvb\ as single photon emitter candidates. In this paper, we employ plane-wave supercell DFT calculations to study the energetic and electronic properties of \vncb\ and \cnvb\ at different charge states and spin multiplicities. We find that the charge states influence the relative stability of the two proposed defects. Although the ZPL energy of \cnvb$(-)$ with quartet spin multiplicity is in the experimental range of interest, the calculated formation energy and charge transition level (CTL) indicate that the quartet state is not stable. Our results imply that the charge states of defects can dramatically affect their relative stability and provide further evidence for the identification of quantum emitters in the visible region. 

\begin{figure}[tb]
\includegraphics[width=\columnwidth]{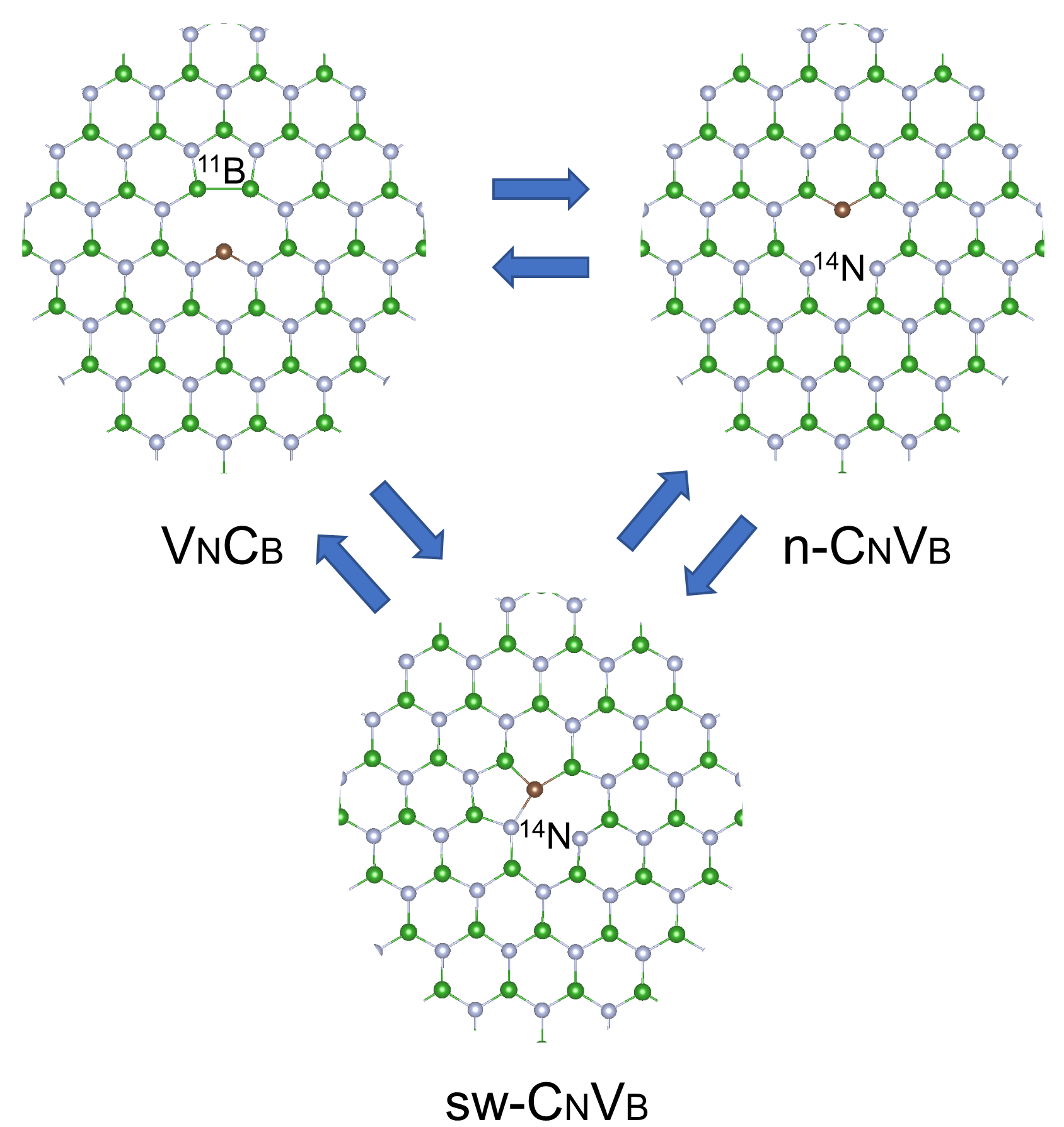}
\caption{\label{Figure 1}%
Schematic view of carbon-vacancy complexes in hBN. The carbon atom can migrate to the antisite-vacancy configuration. Grey, green, and brown colored balls represent nitrogen, boron and carbon atoms, respectively. The usual \cnvb\ considered before is denoted as n-\cnvb.}
\end{figure}

\section{Results}
\label{sec:results}
The ground state electronic structures are calculated for \vncb\ and \cnvb\ in $-1$, $0$ and $+1$ charged states with the high-spin (HS) and low-spin (LS) states included. Previous results about the spin multiplicity of the ground state varied with the applied calculation methods~\citep{sajid2018defect,mendelson2021identifying}. For neutral \vncb$(0)$, Cheng \textit{et al.}~\citep{cheng2017paramagnetic} predicted a triplet ground state, however, further comprehensive calculations indicated that the $^1A_1$ singlet is the most favorable one~\citep{reimers2018understanding}. Our calculation shows that the singlet (LS) state of \vncb$(0)$ has lower energy considering the symmetry reduction caused by the pseudo-Jahn-Teller (JT) distortion in the singlet state while it does not occur in the triplet state. The $C_s$ symmetry configuration is lower in energy by 1.87~eV over that of $C_{2v}$ symmetry configuration. In the following calculation, we remove the symmetry constrain and investigate the electronic structure. We find that the $C_{2v}$ symmetry is the most stable on in the positive charge state regardless the spin multiplicity, and \vncb(+) prefers the low-spin doublet (LS). Meanwhile, \vncb($-$) exhibits doublet state with $C_s$ symmetry.

For the \cnvb\ defect (n-\cnvb\ in this paper), the triplet state is predicted to have lower energy than singlet for \cnvb(0) with transition energy in experimental value~\citep{sajid2018defect}. However, a recent report indicates that this configuration suffers from multiple low-energy minima, and none of them has significant oscillator strength~\citep{mendelson2021identifying}. Noteworthy, we find that the \cnvb\ could form a Stone-Wales-like (sw-\cnvb) configuration, as shown in Fig.~\ref{Figure 1}, which could have lower formation energy. The multireference character still exists here which is manifested as the unrestricted spin-polarized calculation yields different energy for spin-up and spin-down channels. We speculate this phenomenon is due to the spin contamination from the triplet ground state and indeed the calculated energy of the singlet is almost identical to that of the triplet. Although these states are not spin eigenstates, the unrestricted DFT orbitals provide more realistic total energies. Except for HSs of \cnvb($-$), other electronic states prefer the sw-\cnvb\ configuration. In the following discussion, always the most stable configuration is discussed for \cnvb\ defects in the context. LSs always have lower energy whereas the CCSD methods pointed out the HS quartet for \cnvb($-$) could be the ground state~\citep{mendelson2021identifying}. This controversy motivated us to further investigate the relative stability of carbon impurity in these two defects.

\begin{figure*}
\includegraphics[width=1.7\columnwidth]{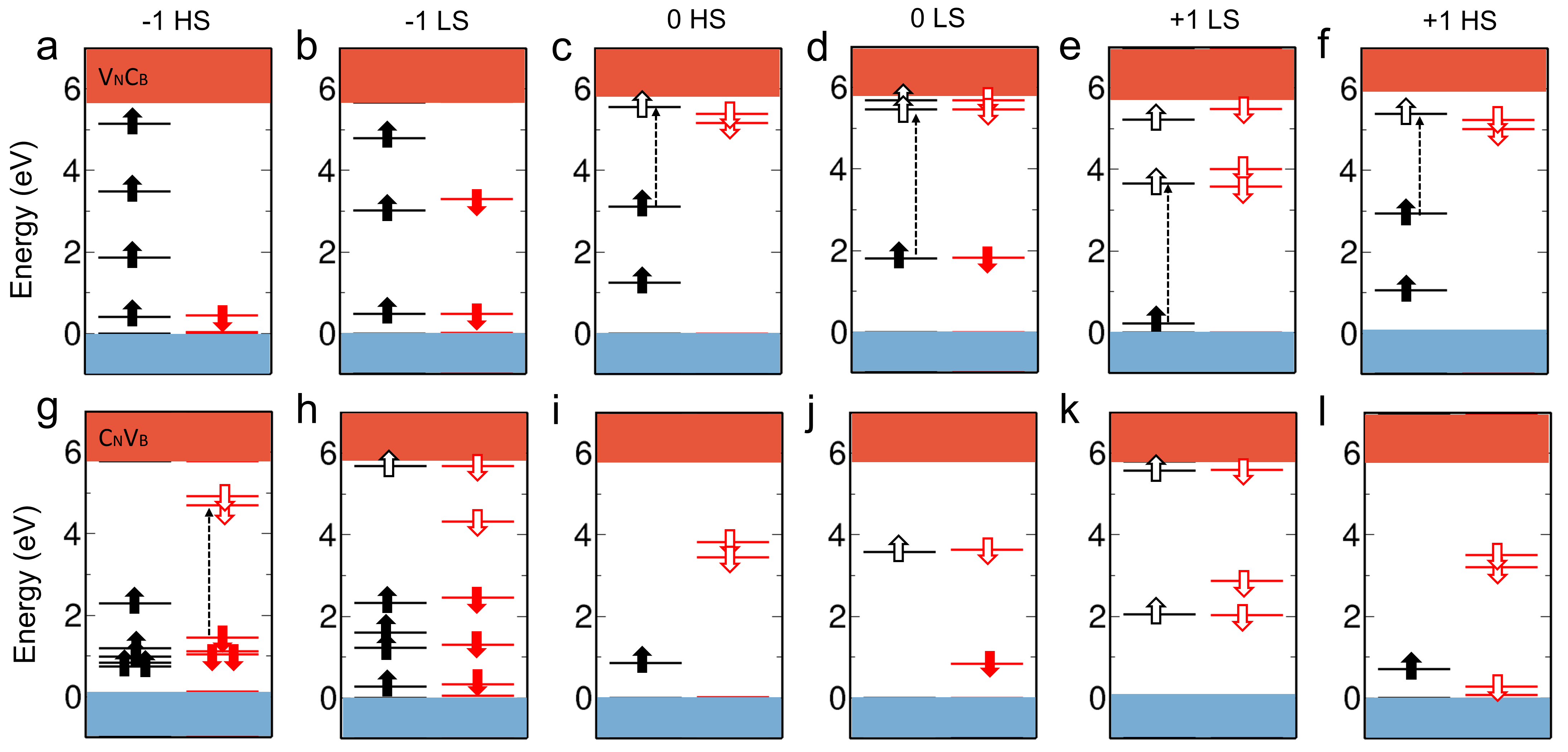}
\caption{\label{Figure 2}%
Energy level diagram of \vncb\ (a-f) and \cnvb\ (g-l) with different charge states and spin multiplicities. The filled and empty arrows indicate the occupied and unoccupied defect states in the spin-up and spin-down channels. Resonant defect levels below valence band maximum are not shown here. The dash lines represent the allowed optical transition in the visible wavelength region. We use the most stable configurations at every charge and spin states. Here, the \cnvb($-2$) and HSs of \cnvb($-$) have n-\cnvb\ configuration and the others are calculated with sw-\cnvb.}
\end{figure*}

The energy level diagrams for the two defects are shown in Fig.~\ref{Figure 2}. \vncb($-$) at both HS and LS states do not have empty localized state therefore only defect-to-conduction band (CB) excitation occurs which results in a relatively dim emission. The experimental observed bright luminescence should come from defect states in the fundamental band gap and neither of the vertical transition energies is in line with the experimental PL energies. The \vncb(0) HS state with $C_{2v}$ symmetry has been studied~\citep{sajid2020vncb}, the defect demonstrates suitable ZPL energy and 1.5 Huang-Rhys (HR) factor. However, the HR factor of the more stable LS state in $C_{s}$ is 7.25, much larger than the experimental value 1.45. The ZPL of \vncb(+) LS is 1.51 eV~\citep{abdi2018color}, which agree relatively well with the experimental data for the Group-2 emitters. Despite of that, the corresponding HR factor is about $24$ so we can also disregard as a good candidate for the observed visible emitters. 
\cnvb($-$) with both HS and LS states have defect-to-defect transitions. The singlet \cnvb(0) has ZPL of 1.98~eV and we believe this value is still in the range of interest even taking into account the possible error due to multireference character of the ground state, however, the HR factor is 8.17. There are no occupied defect states in the gap for \cnvb(+) and the allowed transition is between valence band maximum to defect states which should result in again a relatively dim optical transition.

We summarized the calculated lowest vertical excitation energies in Tab.~\ref{Table 1}. For defects in the negative charge state, the \cnvb($-$) with HS state is the only possible candidate for the bright visible emitters, and the defects in other charge states have either too small transition energy or large phonon side band. The calculated ZPL is 1.89~eV which is close to that obtained in the flake model~\citep{mendelson2021identifying}.

\begin{figure}
    \includegraphics[width=0.8\columnwidth]{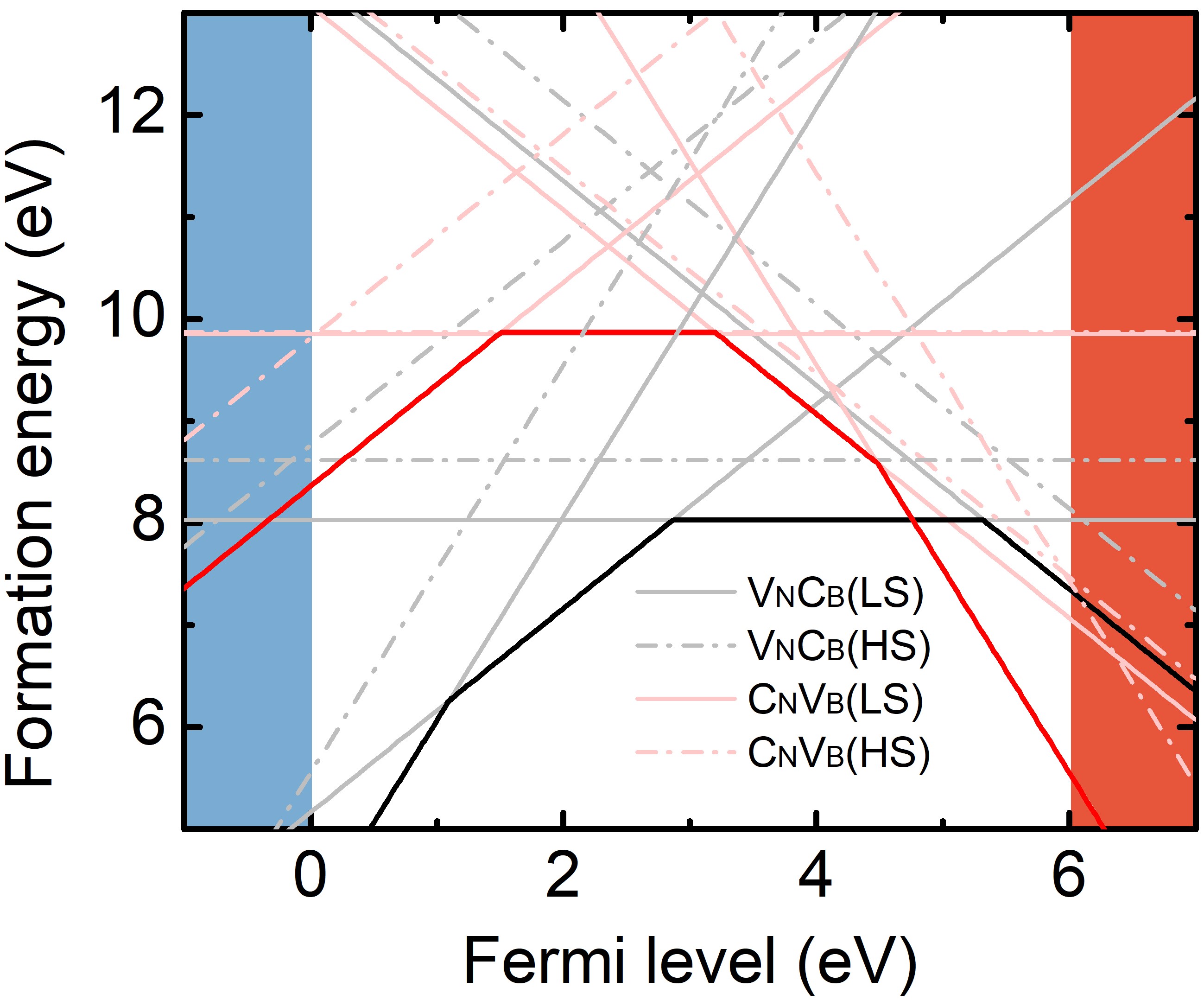}
    \caption{\label{Figure 3}The formation energies of \vncb\ (black) and \cnvb\ (red) as a function of the Fermi-level. The slope of line segment corresponds to the charge state. The solid and dash lines indicate the LS and HS state, respectively. Crossing of the lines denote the charge transition level. Here, the \cnvb($-2$) and HSs of \cnvb($-$) have n-\cnvb\ configuration and the others are calculated with sw-\cnvb.}
\end{figure}

The defect formation energies $E_f$ are calculated to determine the charge stability with the following equation,
\begin{equation}
\begin{split}
E^q_f = &E^q_d - E_\text{per} + \mu_\text{C} -\mu_\text{B} - \mu_\text{N} + q\left(\epsilon^\text{per}_\text{VBM} + \epsilon_\text{Fermi}\right)\\
&+ E_\text{corr}\left(q\right)\text{,}
\end{split}
\end{equation}

where $E_d^q$ is the total energy of hBN model with defect at $q$ charge state and $E_\text{per}$ is the total energy of hBN layer without defect. $\mu_\text{C}$ is the chemical potential of carbon and can be derived from pure graphite. The Fermi-level $\epsilon_\text{Fermi}$ represents the chemical potential of electron reservoir and it is aligned to the valence band maximum (VBM) energy of perfect hBN, $\epsilon^\text{per}_\text{VBM}$. The $E_\text{corr}\left(q\right)$ is the correction term for the charged system due to the existence of electrostatic interactions of the periodic images of the defect. \vncb\ and \cnvb\ always prefers LS state for different charge states. As shown in Fig.~\ref{Figure 3}, in the fundamental gap, the CTL of \vncb\ for $(+1/0)$ is 2.88~eV and it is 5.30~eV for $(0/-1)$ with respect to VBM. These CTLs are far from conduction band minimum (CBM) and VBM, respectively, making \vncb\ a hyper deep donor and acceptor. For \cnvb, the $(+1/0)$ and $(0/-1)$ levels are at 1.51~eV and 3.19~eV, respectively. We notice that the $+2$ charge state of \vncb\ and $-2$ charge state of \cnvb\ are also stable as reported previously~\citep{maciaszek2022thermodynamics}. \vncb($+2$) have $C_{2v}$ symmetry and the \cnvb($-2$) have n-\cnvb\ configuration. However, the calculated CTLs are relatively close to the band edge. $(+2/+1)$ is at 1.11~eV for \vncb\ while the $(-1/-2)$ is at 4.49~eV for \cnvb. Optical excitation results in photoionization to other charge states and thus cannot have optical transition energies in fluorescence close to the experimental values. \vncb\ is generally more stable with lower formation energy than \cnvb\ when $\epsilon_\text{Fermi}$ \textless\ 4.75~eV. The \cnvb($-2$) has lower formation energy with Fermi-level close to CBM which is not a typical experimental condition.

\begin{figure*}
    \includegraphics[width=1.7\columnwidth]{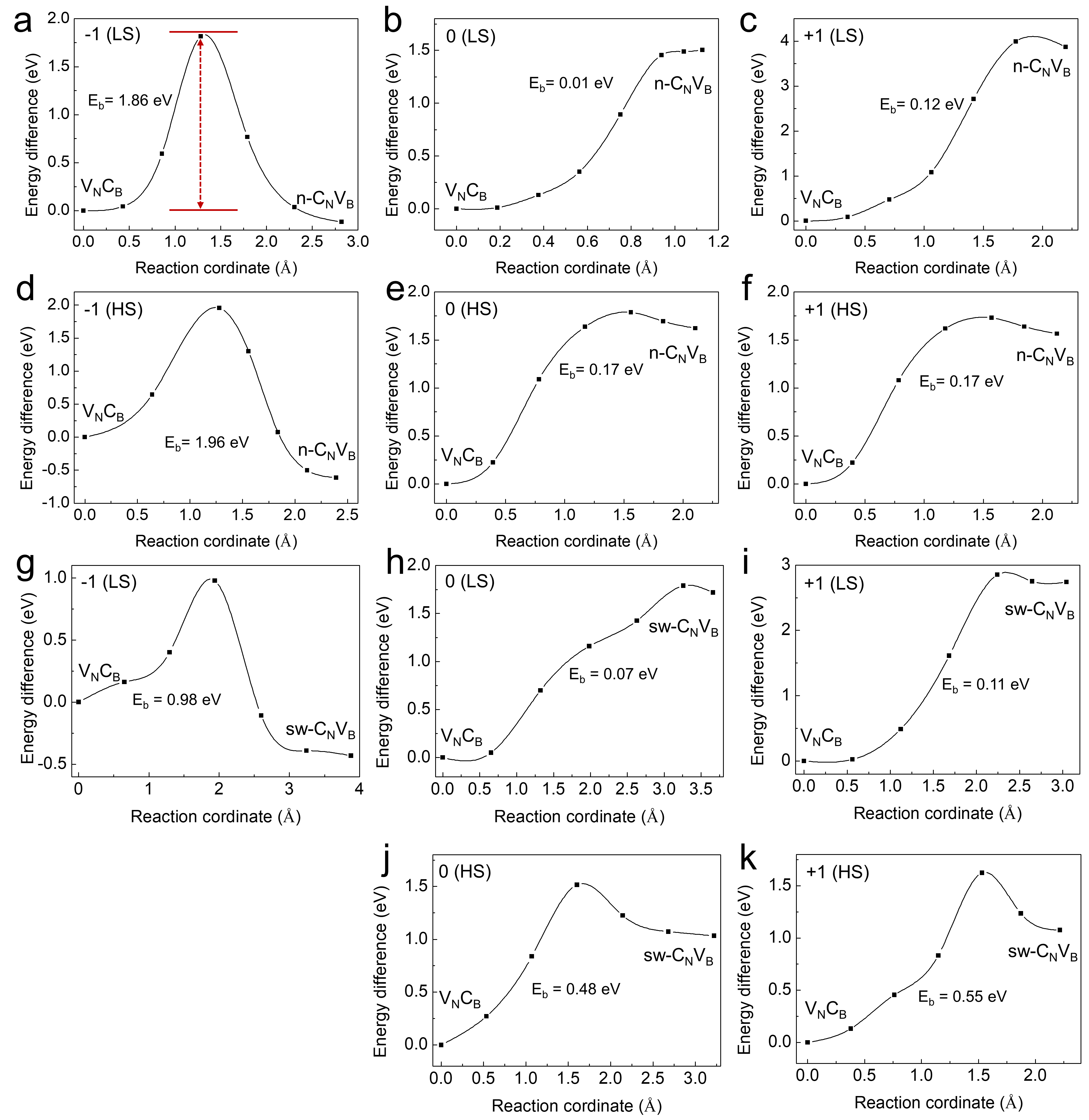}
    \caption{\label{Figure 4}The reaction barrier for the carbon atom from initial state \vncb\ to final state \cnvb\ in different charge states. Transitions from \vncb\ to n-\cnvb\ (a-f) and from \vncb\ to sw-\cnvb (g-k) are shown here. The red arrow indicates the reaction barrier $E_b$. The HSs of \cnvb\ does not have sw-\cnvb\ configuration.}
\end{figure*}

Next, we calculated the defect migration with the climbing-image nudged-elastic-band method~\citep{henkelman2000climbing,henkelman2000improved} as shown in Fig.~\ref{Figure 4}. The n-\cnvb\ and sw-\cnvb\ are both considered here. The carbon atom can migrate between boron site and nitrogen site and this transition rate $\Gamma$ can be simply expressed as~\citep{weston2018native},
\begin{equation}
\begin{split}
\label{eq:migration}
\Gamma = \Gamma_0exp(-\frac{E_b}{k_BT}),
\end{split}
\end{equation}
where $E_b$ is the reaction barrier energy, $k_B$ is the Boltzmann constant, $T$ is the temperature and $\Gamma_0$ is typical phonon frequency in hBN which is about $10^{14}s^{-1}$~\citep{geick1966normal}. The calculated $T$ can be regarded as annealing temperature at which the defect becomes mobile as listed in Tab.~\ref{Table 1}. Usually the temperature is estimated when the jump rate $\Gamma$ is 1/s~\citep{janotti2007native}. Above this temperature, the reaction barrier can be passed to reach an equilibrium point. In the negative charge state, the reaction barrier energies for LSs and HSs are 0.98~eV and 1.96~eV, respectively, for carbon migration from boron site to nitrogen site. The large barrier energies indicate that the carbon atom cannot jump from one site to another, therefore, the two defects could exist simultaneously. The annealing temperature are 353~K and 705~K, therefore, \vncb\ would transform to \cnvb\ if the sample is grown or annealed above these temperatures. However, the barrier energies are quite small in the neutral and positive charge states. The barrier energies are generally less than 0.6~eV, especially, for LSs in the neutral state, manifesting the carbon can move freely from nitrogen site to boron site. The calculated annealing temperature is lower than room temperature, hence the \vncb\ defect will dominate in these two charge states, and the \cnvb\ can only be stabilized through low-temperature irradiation.

The reaction barrier energies and migration paths depend on the charge state. We tentatively associate the low stability of \cnvb(0) and \cnvb(+) to the unoccupied localized states in the gap. Filling these defect levels by electrons could stabilize the defect.

\section{Discussion}
\label{sec:discussion}
We calculated the formation energies and the barrier energy of transformation of carbon-vacancy complexes depending on the symmetry and spin multiplicities.
We note that the geometry distortions from the planar structure was found for certain carbon-vacancy complexes as the common Jahn-Teller distortion effect where the atoms farther the core of the defect structure remain in the sheet of hBN layer. This is different from recent proposed geometry containing dramatically out-of-plane warping calculated by CAM-B3LYP DFT functional~\citep{mendelson2021identifying}. This warping might be induced by the hydrogen termination of the model or the freezing of atoms during optimization. In addition, the large scale distortion might yield opposite result. In our model, the relaxation energy of HS \cnvb($-$) is 0.16~eV and HR factor is 2.2 which are close to the experimental data~\citep{mendelson2021identifying}. However, the formation energy difference between the HS and LS for n-\cnvb($-$) is 0.4~eV, 
where the LS configuration exhibits $C_{2v}$ symmetry. Our NEB calculation shows there is no barrier for n-\cnvb\ to transform to sw-\cnvb\ at LSs for three charge states which indicates the n-\cnvb\ might not exist at all. Recent work indicates that n-\cnvb\ is dynamically unstable and it quickly relax to \vncb\ defect~\citep{ortigoza2022thermodynamic}. This is consistent with our present data for the neutral charge state. We show here that sw-\cnvb\ occurs in the other charge states too, thus this behavior is independent on the position of the Fermi-level of hBN.

Similar to \vb~\citep{gottscholl2020initialization}, hyperfine interaction related features may be a unique fingerprint of the \cnvb($-$) defect observed by electron spin resonance techniques. Recently, an electron paramagnetic resonance (EPR) center has been observed in hBN~\citep{toledo2020identification}. The experimental data clearly shows splitting pattern of five peaks with relative intensities of 1:2:3:2:1. This is due to the hyperfine coupling between one electron spin and two equivalent nitrogen nuclear spin of $I$ = 1, therefore, the EPR center has been tentatively associated with the spin doublet of n-\cnvb($-$) defect~\citep{toledo2020identification}. Although, our calculations indicate that n-\cnvb($-$) does not exist in hBN, we still simulated its EPR spectrum after calculation of the hyperfine tensors of the defect (see Tab.~\ref{Table 2}). Indeed, the simulated EPR spectrum of n-\cnvb($-$) agrees with the experimental one ~\citep{toledo2020identification} (Tab.~\ref{Table 2}). We note that the hyperfine signatures of the HSs configuration significantly differ. The stable sw-\cnvb($-$) has very different spin density distribution with localized on two boron ions, thus its EPR spectrum significantly deviates from that of the observed EPR center. However, n-\cnvb($-$) is not a stable structure, therefore, we conclude that it cannot account for the EPR center as tentatively proposed in Ref.~\citep{toledo2020identification}, despite the agreement between the simulated and observed hyperfine related features in the EPR spectra.

\begin{figure}
    \includegraphics[width=\columnwidth]{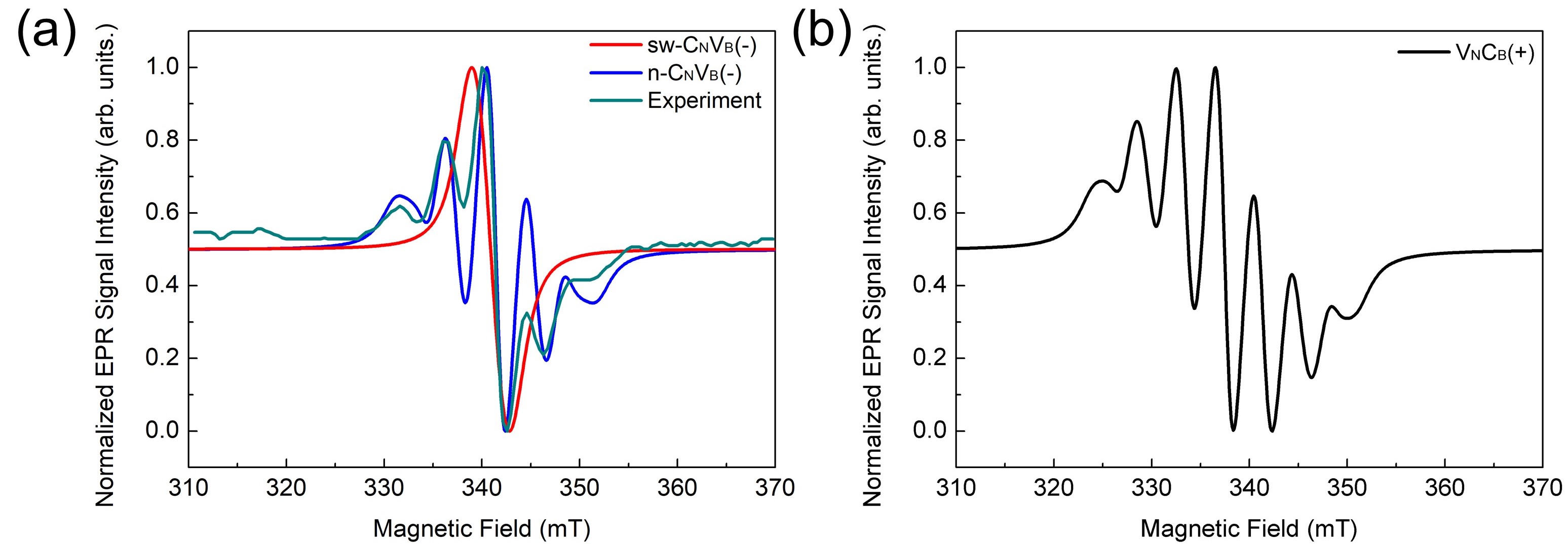}
    \caption{\label{Figure 5}The simulated EPR spectrum of defects considered. (a) The experimental data is from ~\citep{toledo2020identification} recorded at 550 $^{\circ}$C to clearly show the line splitting. sw-\cnvb\ has relative small hyperfine coupling so there is no line splitting. The experimental spectrum is shifted 4 mT towards left to match simulation result. (b) The EPR spectrum of \vncb(+).}
\end{figure}

We found that the most stable form of carbon-vacancy complex in hBN is \vncb\, thus, we plot the simulated EPR spectrum for the paramagnetic \vncb(+) in Fig.~\ref{Figure 5}(b) which may occur in hBN. The EPR spectrum shows 7 peaks due to two neighboring boron atoms ($I=3/2$ for $^{11}$B). Further experimental data is needed to confirm the existence of such kind of defect. 

We note that there is a metastable triplet state for \vncb(0). The triplet state may be accessed by optical pumping of the system from the singlet ground state to the singlet excited state followed by an intersystem crossing from the excited singlet state towards the metastable triplet state. However, the large singlet-triplet energy gap makes this process inefficient, thus, we do not consider the triplet of  \vncb(0) to be observed by photo-EPR studies.

%
%
\section{Summary and conclusion}
\label{sec:summary}
In this study, we performed density functional theory calculations on the carbon-vacancy complexes in hBN. The analyzed relative stability of these complexes could reveal the origin of the single photon emitters observed in experiments in the visible wavelength region. sw-\cnvb\ is always more stable than n-\cnvb\ and n-\cnvb\ transforms to sw-\cnvb\ without any barrier in three charge states. Hence, previously reported n-\cnvb\ is not the origin of visible SPEs in hBN. sw-\cnvb\ is metastable and can transform to \vncb\ with some barrier energies depending on the charge states. In addition, neither the \vncb\ nor \cnvb\ is a potential candidate for the SPEs associated with the carbon impurities. The formation energies for \vncb\ and \cnvb\ complexes imply that these complexes cannot account for the experiments claiming that the number of C-B bond is larger than C-N bond in carbon contaminated hBN~\citep{mendelson2021identifying}. Further investigation is needed for finding the microscopic origin of SPEs associated with the carbon impurities. Recent studies proposed the C$_2$C$_\text{N}$ configuration~\citep{auburger2021towards,li2022carbon,golami2022b} of which optical properties well reproduce the experimental data. We also found that the recently reported EPR center~\citep{toledo2020identification} is not associated with the unstable \vncb\ defect. Further investigation is needed to identify this EPR center in hBN. We provide the EPR spectrum for the most stable carbon-vacancy complex in hBN that might be found in future EPR studies of carbon contaminated hBN layers.

\section{Methods} 
\label{sec:methods}

\renewcommand{\arraystretch}{1.5}
\begin{table}[tb]
\caption{\label{Table 1}  The lowest vertical transition energy for two defects in hBN. The energy unit is eV. The simulated annealing temperature (temp.) is also listed as obtained from the calculated barrier energies (see text). For \cnvb\ defect, except for the \cnvb($-$) HSs, other charge states have lower total energy with sw-\cnvb\ configuration. The $\Delta E_\text{diff}$ indicates energy difference between these two configurations and defined as $\Delta E_\text{diff}$ =  E(n-\cnvb) -  E(sw-\cnvb).}
\begin{ruledtabular}
\begin{tabular}{lcccccc}
configuration & -1 HS & -1 LS & 0 HS & 0 LS & +1 HS & +1 LS  \\
\colrule\rule{0pt}{2.5ex}%
\vncb\ & 0.64 & 0.87 & 1.86 & 1.93 &1.81 & 2.10 \\
SYM. &   $C_{2v}$ & $C_{s}$ & $C_{2v}$ & $C_{s}$ & $C_{2v}$ & $C_{2v}$ \\
HR factor & & & $1.50^\text{a}$ & 7.25 &  & 24\\\hline
n-\cnvb\ & 2.05 & 1.02 & 1.75 & 1.97 &0.81 & 0.91 \\
SYM. &   $C_{2v}$ & $C_{2v}$ & $C_{2v}$ & $C_{2v}$ & $C_{2v}$ & $C_{2v}$ \\
HR factor &2.2 & &  &  &  & \\\hline
sw-\cnvb\ & - & 1.12 & 1.49 & 2.57 &0.14 & 0.49 \\
$\Delta E_{diff}$ & - & 0.37 & 0.70 & 0.90 & 2.0 & 0.78 \\
HR factor & & &  & 8.17 &  & \\\hline
TEMP.(K) & 705 & 353 & 173 & 25 & 198 & 40 \\
\end{tabular}
\end{ruledtabular}
a. Simulated data from Ref.~\cite{sajid2020vncb}
\end{table}
\renewcommand{\arraystretch}{1.0}

\renewcommand{\arraystretch}{1.5}
\begin{table}[tb]
\caption{\label{Table 2}  Hyperfine constants for carbon-vacancy complexes in hBN. The unit is in MHz.}
\begin{ruledtabular}
\begin{tabular}{lcccc}
&configuration & $A_{xx}$ & $A_{yy}$ & $A_{zz}$ \\
\colrule\rule{0pt}{2.5ex}%
$^{14}$N & Exp.~\cite{toledo2020identification} & 90-98 & 90-98 & 145-148  \\
$^{14}$N &n-\cnvb($-$) LS & 86.92 & 83.12 & 166.41 \\
$^{14}$N &n-\cnvb($-$) HS & 34.05 & 33.04 & 59.54  \\
$^{14}$N &sw-\cnvb($-$) LS & 6.75 & 6.53 & 13.25
  \\\hline
$^{11}$B &\vncb($+$) LS & 101.47 & 100.75 & 126.87 \\
\end{tabular}
\end{ruledtabular}
\end{table}
\renewcommand{\arraystretch}{1.0}

We performed spin-polarized DFT calculation within the Kohn-Sham scheme as implemented in the VASP code~\citep{kresse1996efficiency,kresse1996efficient}. Standard projector augmented wave (PAW) formalism ~\citep{blochl1994projector,kresse1999ultrasoft} is used to separate the valence electrons from nuclei. The convergence threshold is 0.01~eV/\AA\ for force acting on each atoms and energy cutoff for the expansion of the plane-wave basis set is 450~eV. The screened hybrid density functional of Heyd, Scuseria, and Ernzerhof (HSE) ~\citep{heyd2003hybrid} is used to calculate the electronic structure and localize bound states. In this approach, we could mix part of nonlocal Hartree–Fock exchange to the generalized gradient approximation of Perdew, Burke, and Ernzerhof (PBE) with fraction $\alpha$. $\alpha$ = 0.32 can reproduce the experimental band gap about 6~eV. We embedded the carbon defects in a $9\times5\sqrt{3}$ monolayer supercell with 162 atoms which is sufficient to avoid the periodic defect-defect interaction, a vacuum layer of 12 \AA\ is applied to separate the periodic layer images. The single $\Gamma$-point scheme is converged for the k-point sampling for the Brillouin zone. The excited states were calculated by $\Delta$SCF method~\citep{gali2009theory}.
For the formation energy calculation, a bulk model with two hBN layers is used to include the interlayer interaction and decreases the artificial influence induced by vacuum layer. The charge correction term is computed by SXDEFECTALIGN code from Freysoldt method~\citep{freysoldt2018first}. During the NEB calculation, for \vncb\ to \cnvb\ transition, the threshold of force is set to 0.02~eV/\AA\ whereas for n-\cnvb\ to sw-\cnvb\ transition it is 0.1~eV/\AA\ due to the numerical challenges. The electron paramagnetic resonance (EPR) simulation is performed with EASYSPIN software at X band region (9.45~GHz)~\citep{stoll2006easyspin}. 

The total HR factor is defined as the number of an effective phonon participating in the optical transition which is a key parameter of the absorption and fluorescence spectra. The total HR factor can be readily calculated within Franck-Condon approximation which assumes that the vibrational modes in the ground and excited states are identical. The associated phonon overlap spectral function can be derived from the overlap between the phonon modes in the electronic ground and excited states~\citep{alkauskas2014first,gali2009theory}.   

\section*{Author contribution}
All authors contributed to the discussion and writing the manuscript. AG led the entire scientific project.

\section*{Competing interests}
The authors declare that there are no competing interests.

\section*{Data Availability}
The data that support the findings of this study are available from the corresponding author upon reasonable request.

%
%
\begin{acknowledgments}
AG acknowledges the Hungarian NKFIH grant No.~KKP129866 of the National Excellence Program of Quantum-coherent materials project and the support for the Quantum Information National Laboratory from the Ministry of Innovation and Technology of Hungary.
\end{acknowledgments}

\bibliography{mainref}

\end{document}